\documentclass{article}

\usepackage{arxiv}

\usepackage[utf8]{inputenc} % allow utf-8 input
\usepackage[T1]{fontenc}    % use 8-bit T1 fonts
\usepackage{hyperref}       % hyperlinks
\usepackage{url}            % simple URL typesetting
\usepackage{booktabs}       % professional-quality tables
\usepackage{amsfonts}       % blackboard math symbols
\usepackage{nicefrac}       % compact symbols for 1/2, etc.
\usepackage{microtype}      % microtypography
\usepackage{graphicx}
\usepackage{multicol}

\title{FitsGeo --- Python package for PHITS geometry development and visualization}

\author{ \href{https://orcid.org/0000-0003-3120-3401}{\includegraphics[scale=0.06]{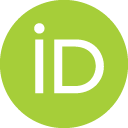}\hspace{1mm}Ivan Gordeev}\\
	Joint Institute for Nuclear Research\\
	Dubna, Moscow region, 141980, Russia \\
	\texttt{gordeev@jinr.ru} \\
}

\hypersetup{
pdftitle={FitsGeo --- Python package for PHITS geometry development and visualization},
pdfsubject={q-bio.NC, q-bio.QM},
pdfauthor={Ivan Gordeev},
pdfkeywords={Python, VPython, PHITS, Geometry Setup, Monte Carlo, Particle Transport},
}

\begin{document}
\maketitle

\begin{abstract}
An easy way to define and visualize geometry for PHITS input files introduced. Suggested FitsGeo Python package helps to define surfaces as Python objects and manipulate them conveniently. VPython assists to view defined geometry interactively which boosts geometry development and helps with complicated cases. Every class that sets the surface object has methods with some extra properties. As well as geometry generation for PHITS input, additional modules developed for material and cell definition. Any user with a very basic knowledge of Python can define the geometry in a convenient way and use it in further research related to particle transport.
\end{abstract}

% keywords can be removed
\keywords{Python \and VPython \and PHITS \and Geometry Setup \and Monte Carlo Modeling \and Particle Transport}

\begin{multicols}{2}
\section{Motivation and significance}
\label{sec:motiv}

Every Monte Carlo (MC) transport code such as MCNP~\cite{MCNP}, FLUKA~\cite{FLUKA} or PHITS~\cite{PHITS} have a part with geometry construction. This part of work with MC codes usually is the most time-consuming and require some concentration from user. Usually, the geometry preparation represents writing some lines (commands) with specific for transport code formatting in so-called ``input file''. Full process of geometry setup can be defined in 3 steps:
\begin{enumerate}
\item Materials definition.
\item Surfaces definition.
\item Cells definition.
\end{enumerate}

Each step has its special ``section'' in the input file. On the first step, the materials section must be defined, which describes certain materials (e.g. water, void and so on) for future geometry objects. In the second step, the user defines basic primitive objects --- surfaces, such as planes, spheres, boxes, cones and so on. Defined surfaces later can be combined together in some way to represent ``cell'' (in MCNP and PHITS) --- more concrete physical volume, with certain material, which also defined in other special section on the third step of geometry setup. Through these cells transport of particles will go during program execution. And in these cells some physical quantities (energy deposition, flux, yield, etc) of radiation scored in so-called ``tallies'' (another special section).

Visualization of the geometry construction process is an extremely necessary thing, which is somehow provided by the MC codes. One of them, for example, FLUKA, provides special graphical user interface (GUI) --- Flair~\cite{Flair} written in Python. Flair is a powerful tool for creating input files. In this interface, the user has an almost pure CAD system (based on the ``flair-geoviewer'' C++ library), so the part of the work related to configuring the geometry setup becomes easier, since every defined object is displayed in 3D and can be changed interactively. This 3D visualization helps to avoid mistakes and typos when configuring the geometry.

\begin{figure}[htb]
\center
\includegraphics[width=0.8\linewidth]{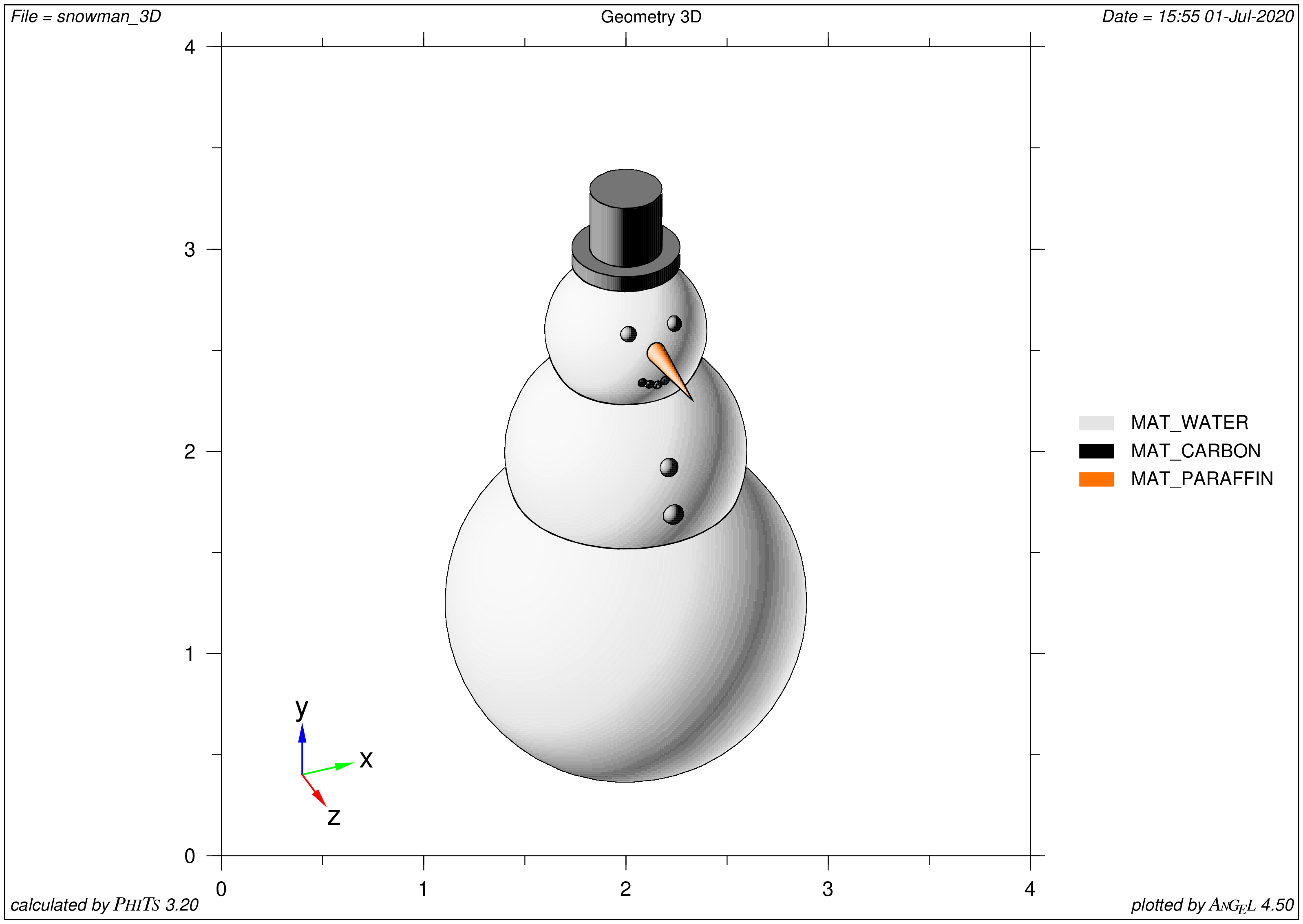}
\caption{The ``Snowman'' example. PHITS geometry visualization via ANGEL}
\label{img:snowman_3D}
\end{figure}

Other codes, such as PHITS, have only limited features for visualization. PHITS uses ANGEL software that creates EPS (stands for Enhanced PostScript) files from PHITS input file. This means that the user cannot get a pure interactive 3D image of the created geometry and for visualization purposes only has eps files where the geometry is displayed from one perspective (see Figure~\ref{img:snowman_3D} for an example of such visualization). In other words, every time a user wants to see what his current geometry looks like from all sides, one need to change the ``camera'' settings, re-launch PHITS in 3D creation mode, and look at the generated eps image, for another perspective other settings are needed, and this process can take some time and cause inconvenience. And this visualization is not interactive (no zooming, no rotating, no panning), so the user can't just ``fly'' through the geometry setup and look at each side of created geometry.
\end{multicols}

\begin{multicols}{2}
Also, if one need to create a really complex geometry with a lot of objects, it is just very difficult to remember all the parameters and set them accordingly by hand every time, not to mention appearing typos while this process. Though, in PHITS user can specify variables, but this helps only partially.

These problems with poor visualization and quite complicated geometry setup by hand encouraged to make a tool that makes this specific part of work with geometries easier and more illustrative. FitsGeo can help the user focus more on a concrete geometry by providing high-quality 3D visualization using VPython and automated generation of geometry objects to PHITS understandable format. The object-oriented programming (OOP) paradigm is a powerful way to prepare such input files, which opens up new possibilities for configuring geometry.

FitsGeo can be very helpful for all researchers working in the field of particle physics, shielding, accelerator design, radiation safety, and other fields of applied physics in which MC transport codes (so far only PHITS) are a common tool. Furthermore, some additional features are provided with the definition of other sections (material and cell sections). The object-oriented way to define geometry and other sections of input files has much more possibilities and provides more flexibility for the user. In addition, this package can be the basis for future development of the GUI for PHITS.

\section{Software description}
\label{sec:descroption}

FitsGeo\footnote{GitHub repository of project --- https://github.com/GordoNice/fitsgeo.git} is a Python package with a set of modules. Each module provides a specific set of classes and some additional functions that a user with very basic Python knowledge can use to create a Python script file for configuring geometry. The object-oriented way of creation of geometry surfaces and additional sections for PHITS input files means that all defined surfaces, materials, and cells are Python objects with some additional properties (get and set methods) such as volume, surface area, density and so on. Thus, the user can change these objects ``on the fly'' when coding geometry. Each object has a number of getting and setting methods, which provides more flexibility for the user.

\newpage

Modules in FitsGeo Package:
\begin{itemize}
  \item \textit{material} module handles material definitions: materials for geometries can be set from predefined databases or manually;
  \item \textit{const} module consists of constants used in FitsGeo: colors for surfaces as VPython vectors and ANGEL colors associated to these colors (in the Python dictionary);
  \item \textit{surface} module consists of classes for defining surfaces --- this is the main class which provides visualization functionallily of FitsGeo;
  \item \textit{cell} module consists of class to define cells;
  \item \textit{export} module provides functionality for exporting all defined objects to the PHITS understandable format (other MC codes may be added in the future releases of FitsGeo).
\end{itemize}

Additional modules for other sections of PHITS input, such as the source definition module will come soon.

Being a Python package, FitsGeo requires Python interpreter. To work with FitsGeo user need to install it first (via pip tool), create Python script file and import FitsGeo package in the script. After these simple steps, full functionality of FitsGeo is available. From now on, every aspect with geometry setup can be set as Python object.

\section{Illustrative Example}
\label{sec:example}

\begin{figure}[h!tb]
\center
\includegraphics[width=\linewidth]{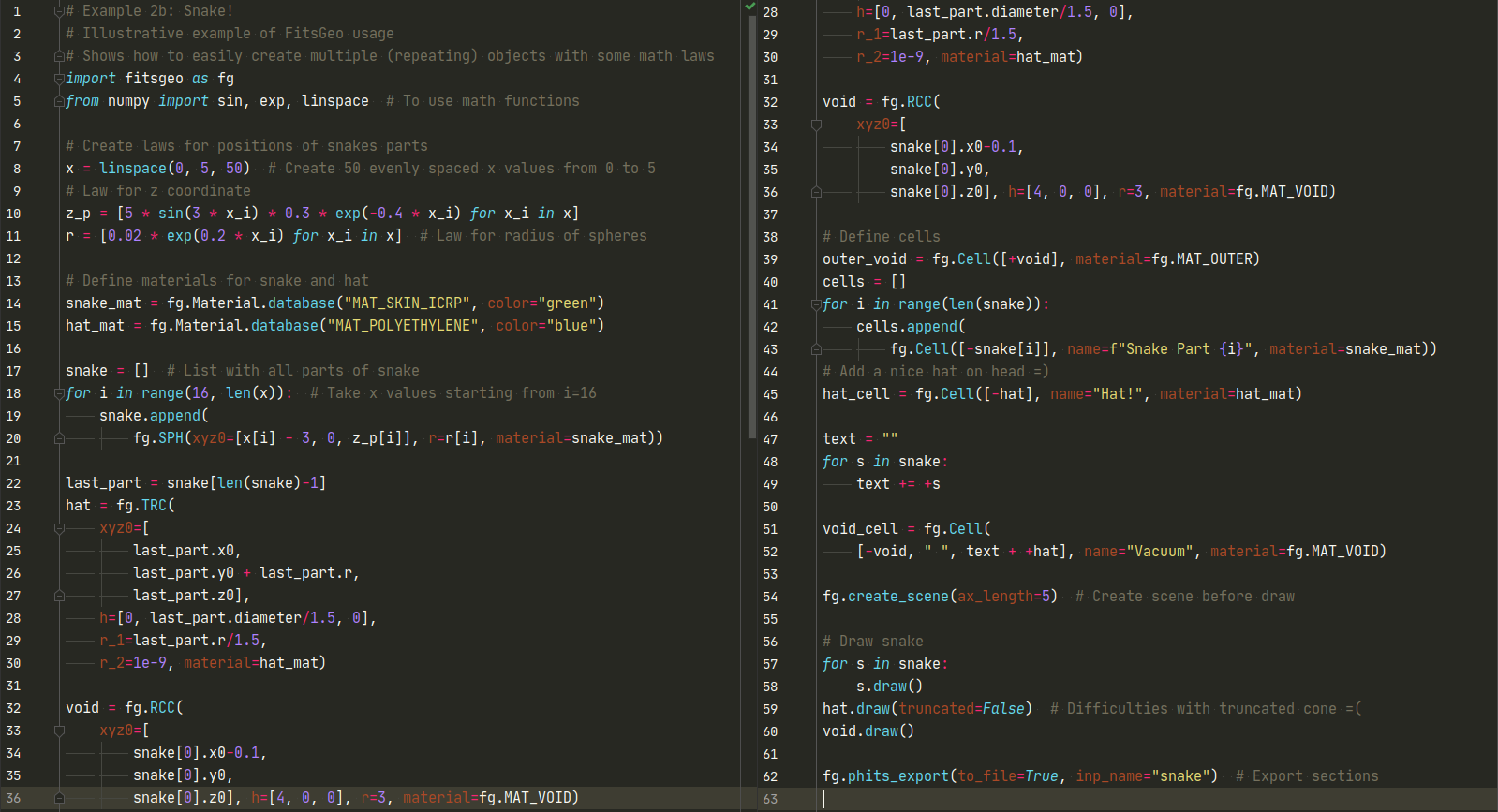}
\caption{The full code of FitsGeo Example 2(b): Snake!}
\label{img:snake_code}
\end{figure}

For the full user guide and examples of using FitsGeo, see the FitsGeo documentation\footnote{FitsGeo documentation website --- https://fitsgeo.readthedocs.io}. Only one example is discussed here, called ``Snake'' (Example 2(b): Snake! on the website). The full code for this example is shown in the Figure~\ref{img:snake_code}.
\end{multicols}

\begin{multicols}{2}
This example demonstrates a workflow for creating multiple (repeating) objects whose placement and size depend on certain mathematical laws. The geometry consists of sphere-segments of the snake's body. Each part of the snake's body defined according to certain math laws for position and size. The example can be divided into 5 parts:
\begin{enumerate}
  \item Definition of math laws for positioning and sizing of snake's parts (lines \textbf{7--11}).
  \item Definition of materials (lines \textbf{13--15}).
  \item Definition of surfaces (lines \textbf{17--36}).
  \item Definition of cells (lines \textbf{38--52}).
  \item Geometry draw and export of all defined objects to the PHITS input file as specific sections (lines \textbf{54--62}).
\end{enumerate}

Before the actual definition of geometry, the user must import modules (lines \textbf{4--5}). In line \textbf{4}, FitsGeo package imported with \textit{fg} alias. Additional functions from the NumPy package are imported for mathematical expressions.

The first part of the code in the above list uses the NumPy module to set mathematical laws of positioning and sizing of snake segments. Let's assume that this snake consists of sphere-segments that must be arranged according to some mathematical law. Assume the following equation for the positions of the sphere-segments on the $xz$-plane:

\begin{equation}
z(x) = 5\cdot\sin(3x)\cdot0.3\exp(-0.4x)
\label{eq:space}
\end{equation}

\begin{figure}[h!tb]
\center
\includegraphics[width=0.7\linewidth]{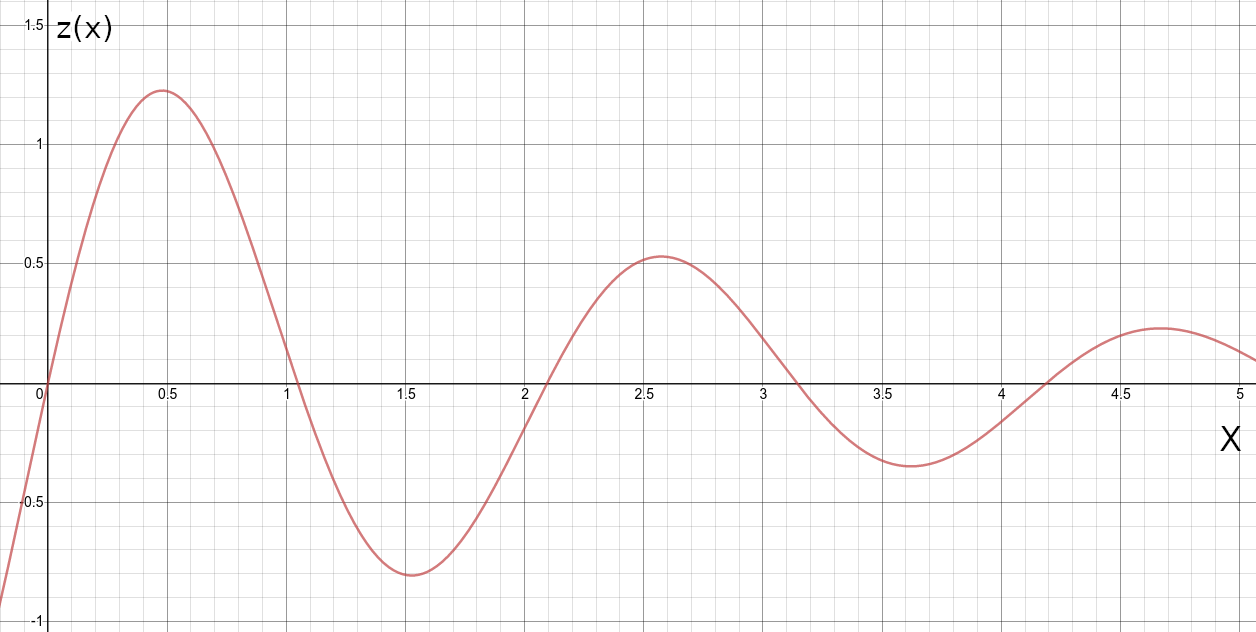}
\caption{Mathematical law for the positions of sphere-segments of a snake}
\label{img:graph_pos}
\end{figure}

This is a formula for $z$ coordinates, where for $x$ coordinates we take 50 evenly spaced numbers from 0 to 5. And on the $y$ coordinate it is 0 everywhere. The graph for this equation is in the Figure~\ref{img:graph_pos}.
\end{multicols}

\begin{multicols}{2}
The equation for the radii of the sphere-segments:

\begin{equation}
r(x) = 0.02\exp(0.2x)
\label{eq:radius}
\end{equation}

\begin{figure}[h!tb]
\center
\includegraphics[width=0.7\linewidth]{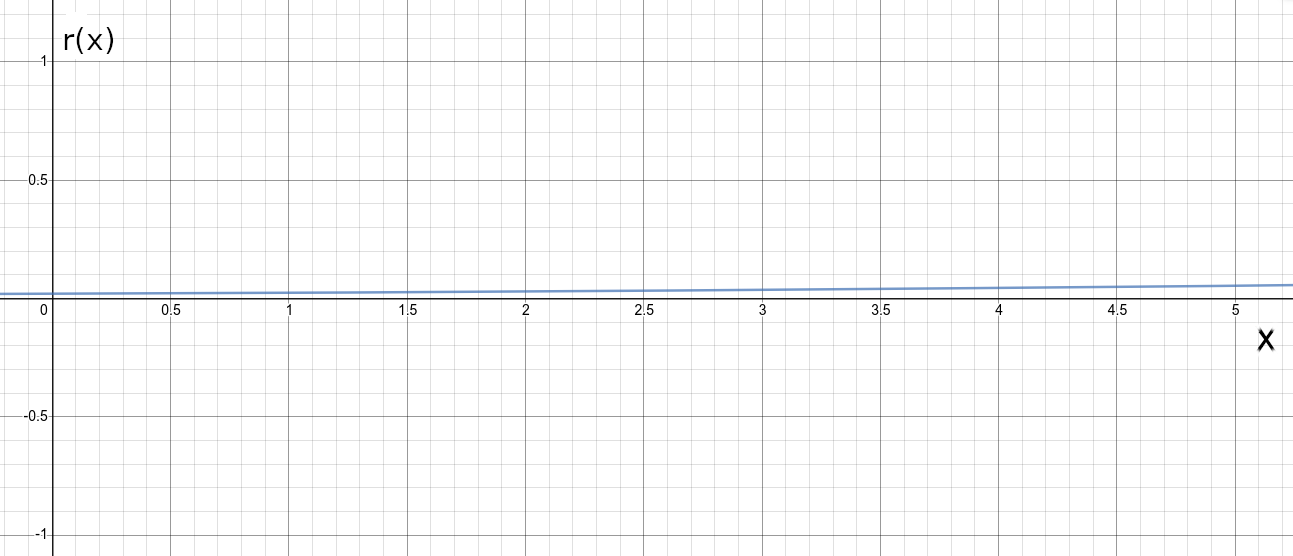}
\caption{Mathematical law for the size of the radii of sphere-segments of a snake}
\label{img:graph_r}
\end{figure}

This law will give us a slow exponential (almost linear) increase in the radius of the sphere-segments depending on the $x$ coordinate. The graph for this law is shown in the Figure~\ref{img:graph_r}.
\end{multicols}

\begin{multicols}{2}
Using FitsGeo, sphere surface objects are placed on points according to the law from the Equation~(\ref{eq:space}) and with radii according to the law from the Equation~(\ref{eq:radius}). This shows why an object-oriented way of defining surfaces can be useful for such geometries.

The second part of the code defines materials for the geometry from predefined databases. FitsGeo has databases with more than 500 predefined materials adapted from GEANT4\footnote{Official website of Geant4 Collaboration --- http://cern.ch/geant4} and SRIM\footnote{SRIM official website --- http://srim.org/} MC codes. In this example, the ``ICRP skin'' and ``Polyethylene'' materials were initialized.

The third part of the code creates geometry surfaces (sphere-segments of the snake and a hat on the head of the snake). Here, the flexibility of FitsGeo becomes more clear, because the user can place surface objects relative to other objects, so the user can even place a hat on the snake's head. It is noticeable that such manipulations is unlikely to be achieved without FitsGeo.

The fourth part of the code is responsible for defining cells. Here we see another advantage: the user can define combinations of surfaces using their names, rather than numbers, as is the case in PHITS. Cells are defined by treating regions divided by surfaces. Surface classes in FitsGeo have overloaded Python operators ``+'' (\_\_pos\_\_) and ``-'' (\_\_neg\_\_), which makes it possible to define ``surface sense''. These operators return the numbers of the surface objects as strings.

\begin{figure}[htb]
\center
\includegraphics[width=0.55\linewidth]{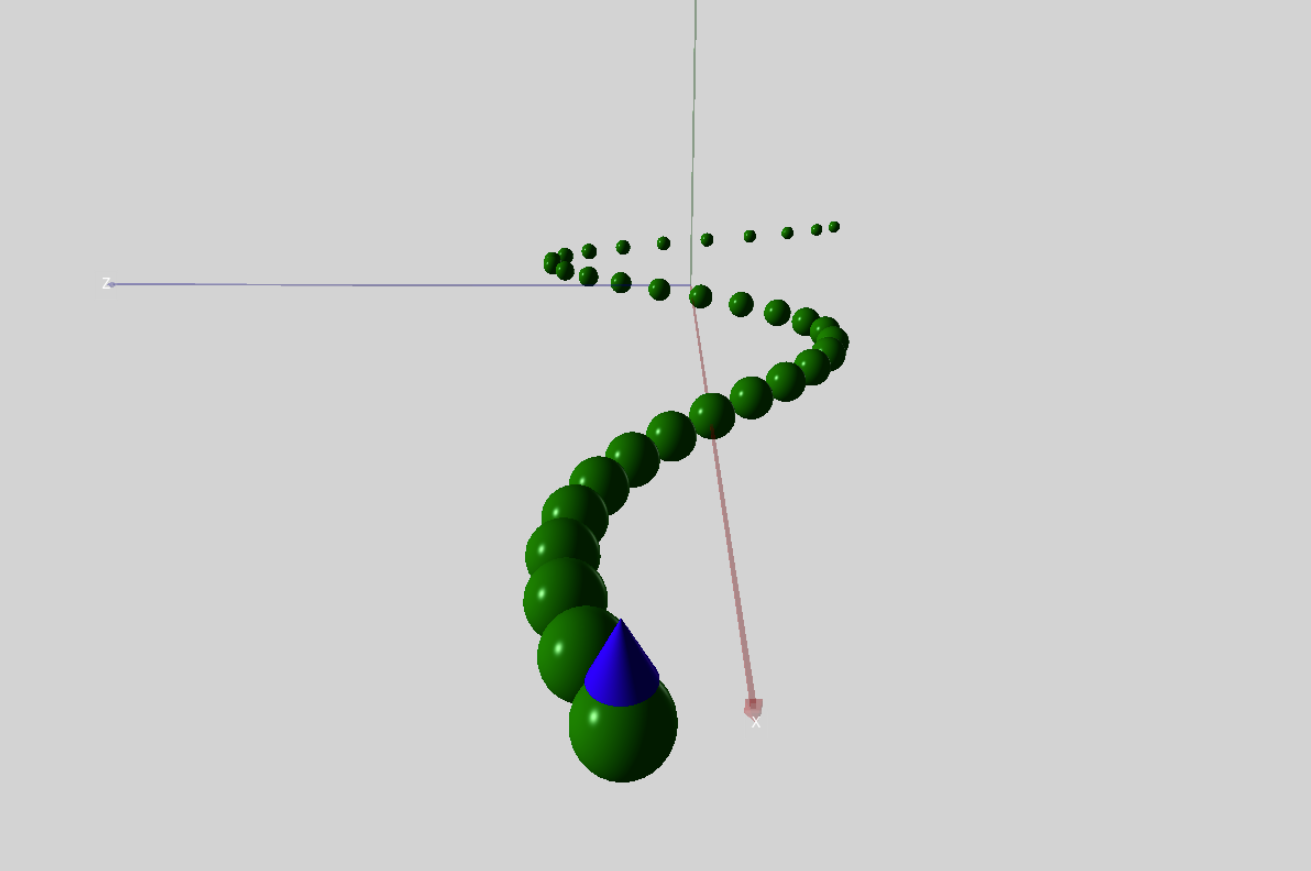}
\caption{Geometry visualization of FitsGeo Example 2(b): Snake!}
\label{img:snake_fg}
\end{figure}

Finally, the fifth part of the code is responsible for the visualization and exporting. In each object method, the user can pass additional parameters (flags) for optional additional functionality, such as changing the opacity of the surface, putting labels that point to the center of the surface, or exporting only certain sections to the PHITS input file, and so on. After executing the ``draw'' method of the surface, the browser tab opens automatically, and the user sees the created geometry in 3D and can interactively view it. This visualization is based on the VPython. The screenshot of such visualization is in the Figure~\ref{img:snake_fg}.
\end{multicols}

\begin{figure}[h!tb]
\center
\includegraphics[width=0.8\linewidth]{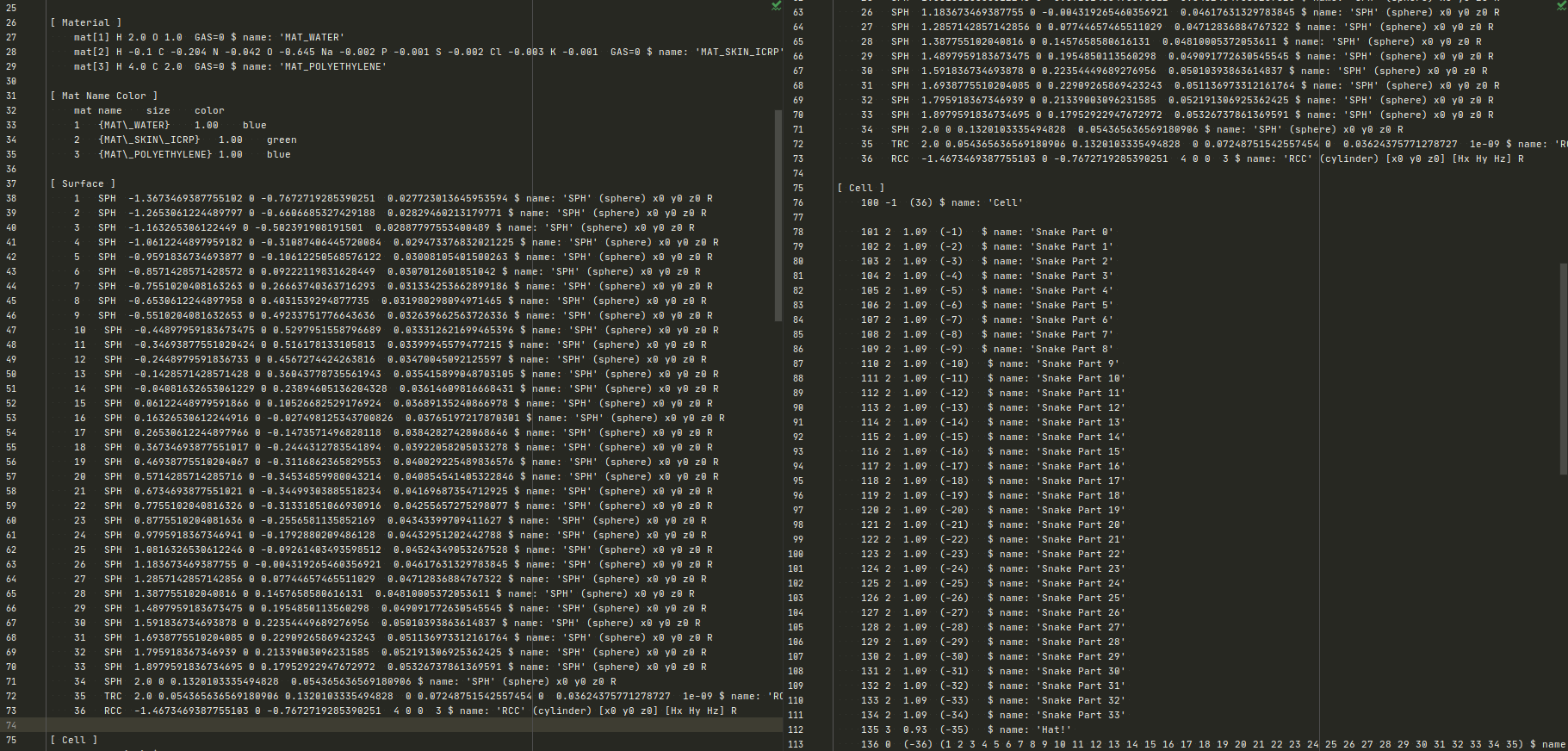}
\caption{The exported from FitsGeo sections for the FitsGeo Example 2(b): Snake!}
\label{img:snake_phits}
\end{figure}

\begin{figure}[h!tb]
\center
\includegraphics[width=0.8\linewidth]{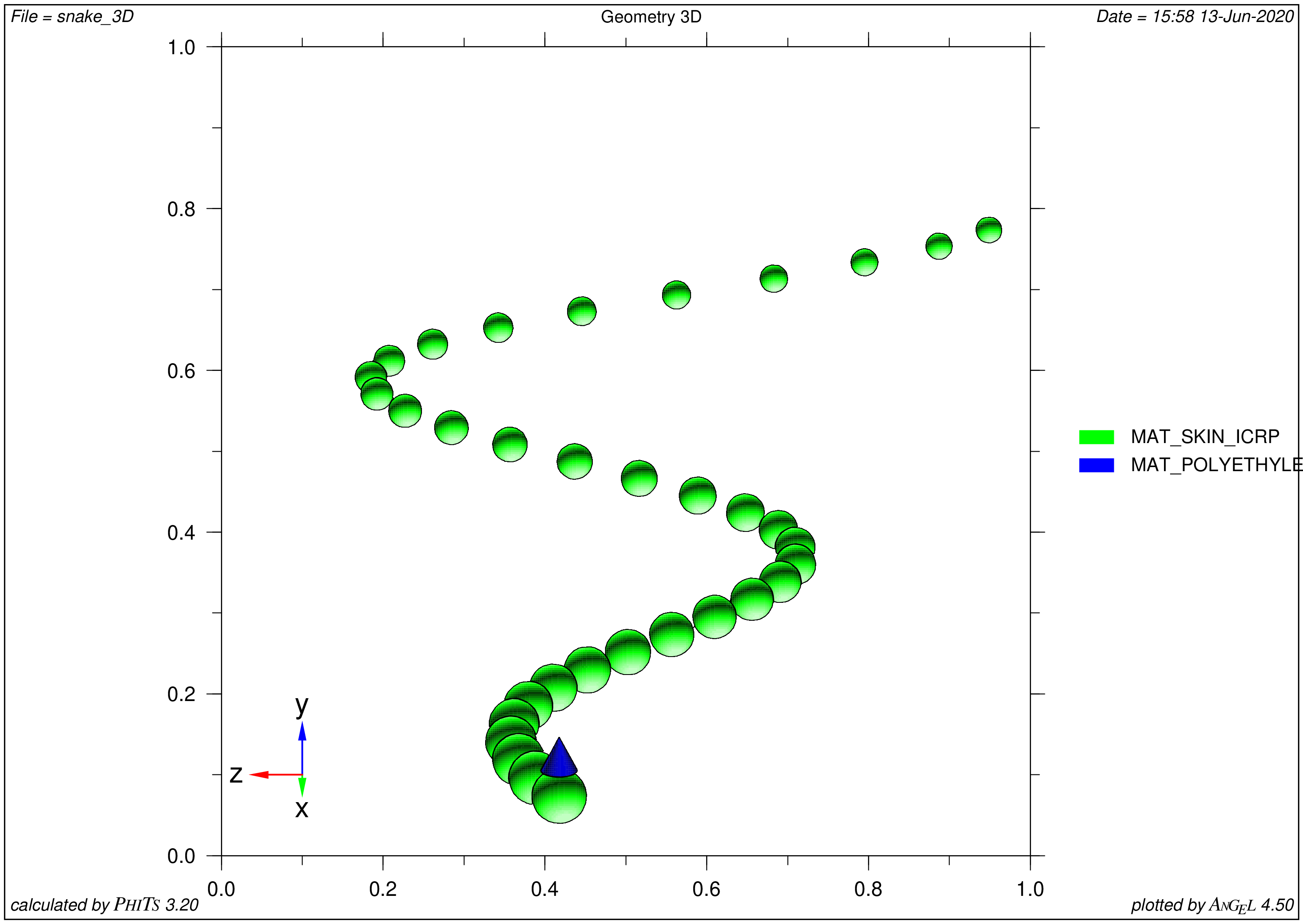}
\caption{Standard visualization of the ``Snake'' example in PHITS via ANGEL}
\label{img:snake_3D}
\end{figure}

\begin{multicols}{2}
The sections exported to PHITS are shown in the Figure~\ref{img:snake_phits}. This is generated automatically, obviously it is very difficult to create such a geometry manually using only the standard way. After export of these sections to the input file, user can perform standard visualization in PHITS via ANGEL (see Figure~\ref{img:snake_3D}).
\end{multicols}

\begin{multicols}{2}
\section{Impact}
\label{sec:impact}

A new way to configure the geometry for PHITS input files has been introduced. The described way has certain advantages, such as an object-oriented way for determining surfaces, materials and cells, improved interactive visualization (using VPython), and automatic export of defined sections to the PHITS input file.

The main features of FitsGeo:
\begin{itemize}
\item Great visualization capabilities with VPython.
\item Easy geometry setup.
\item Python and the OOP paradigm provide more flexibility for geometry development.
\item Additional properties for each type of defined surface (volume, surface area, etc).
\item Databases with 500+ predefined materials.
\item Export of defined objects as sections of the PHITS input file.
\item FitsGeo is well documented and simple even for a beginner in Python.
\end{itemize}

The tool is suggested to be used for a certain group of researchers engaged in MC modeling of particle transport. The described features will certainly improve geometry development for researchers working with PHITS (export to other codes may be done in future releases). With this package, researchers can be more focused on the most time-consuming part of work with PHITS. This will lead to faster development and less mistakes during the geometry configuration stage.

\section{Conclusions}
\label{sec:concl}

The ``FitsGeo'' Python package is presented. This package can be a really useful tool for those who works with PHITS code and need more degree of freedom in developing geometry and visualization. A set of modules with classes and functions make it easy to configure the geometry. Surfaces, materials, cells can be set up as Python objects with all the advantages that follow, they are easy to visualize, and the user can easily export them to a format that is understandable for PHITS. The FitsGeo package is well documented and has a bunch of usage examples, so even a user who is far from programming can easily start developing the geometry and use it for further research. 

\bibliographystyle{unsrt}
% \bibliography{cites.bib} %%% Remove comment to use the external .bib

\newpage


\end{multicols}

\end{document}